\documentclass{article}

\usepackage{PRIMEarxiv}

\errorcontextlines\maxdimen

\usepackage[utf8]{inputenc} 
\usepackage[T1]{fontenc}    
\usepackage{url}            
\usepackage{booktabs}       
\usepackage{nicefrac}       
\usepackage{microtype}      
\usepackage{fancyhdr}       
\usepackage{graphicx}
\usepackage{xcolor}
\usepackage{threeparttable}
\usepackage{booktabs}
\usepackage{multirow}
\usepackage{tabularx}
\usepackage{arydshln}
\usepackage{subcaption}
\usepackage{cite}

\usepackage{lineno,hyperref}
\hypersetup{colorlinks, citecolor=blue, linkcolor=red, urlcolor=blue}
\usepackage{amsmath,amsfonts,amssymb}
\usepackage{graphics}
\usepackage{bm}
\usepackage{multirow}
\usepackage{algorithm}
\usepackage{algpseudocode}

\usepackage{mathtools}
\usepackage{siunitx}
\DeclarePairedDelimiterXPP\BigOSI[2]%
  {\mathcal{O}}{(}{)}{}%
  {\SI{#1}{#2}}
\newcommand{\bd}{\boldsymbol}

\title{A dual physics-informed neural network for topology optimization}

\author{Ajendra Singh \\
  Department of Civil Engineering \\
  Indian Institute of Technology Roorkee\\
  \texttt{a\_singh4@ce.iitr.ac.in} \\
  \And
  Souvik Chakraborty \\
  Department of Applied Mechanics\\
  Yardi School of Artificial Intelligence (ScAI)\\
  Indian Institute of Technology Delhi\\
  \texttt{souvik@am.iitd.ac.in} \\
  \And
  Rajib Chowdhury \\
  Department of Civil Engineering\\
  Indian Institute of Technology Roorkee\\
  \texttt{rajib.chowdhury@ce.iitr.ac.in} \\
}

\begin{document}

\maketitle

\begin{abstract}
We propose a novel dual physics-informed neural network for topology optimization (DPNN-TO), which merges physics-informed neural networks (PINNs) with the traditional SIMP-based topology optimization (TO) algorithm. This approach leverages two interlinked neural networks—a displacement network and an implicit density network—connected through an energy-minimization-based loss function derived from the variational principles of the governing equations. By embedding deep learning within the physical constraints of the problem, DPNN-TO eliminates the need for large-scale data and analytical sensitivity analysis, addressing key limitations of traditional methods. The framework efficiently minimizes compliance through energy-based objectives while enforcing volume fraction constraints, producing high-resolution designs for both 2D and 3D optimization problems. Extensive numerical validation demonstrates that DPNN-TO outperforms conventional methods, solving complex structural optimization scenarios with greater flexibility and computational efficiency, while addressing challenges such as multiple load cases and three-dimensional problems without compromising accuracy.
\end{abstract}


\section{Introduction}
\label{sec:introduction}
Structural topology optimization (TO) plays a pivotal role in designing high-performance, lightweight structures by determining the optimal material distribution within a specified design space, all while adhering to engineering constraints. Although TO has evolved significantly since the foundational work of Bendsoe and Kikuchi \cite{bendsoe1988generating}, there is a pressing need to address certain limitations in current methodologies. Traditional TO approaches, such as density-based methods like the Solid Isotropic Material with Penalization (SIMP) \cite{bendsoe1999material, sigmund200199} and evolutionary methods \cite{xie1993simple, querin1998evolutionary}, have demonstrated success in diverse industries including aerospace, automotive, and architecture sectors. However, these methods often struggle with issues such as convergence speed and computational cost. Newer approaches, such as level-set methods \cite{wang2003level, allaire2004structural}, topological sensitivity methods \cite{mirzendehdel2015pareto, deng2015multi}, and moving morphable components \cite{du2019moving, guo2014topology, zhang2017structural}, offer alternatives but are also computationally expensive. 
As industries demand increasingly sophisticated, multi-functional designs, there is an unmet need for 
developing efficient yet accurate topology optimization algorithms. This paper attempts to address this apparent gap by developing a dual physics-informed neural network for topology optimization, which leverages the power of machine learning to enhance traditional methods and can potentially facilitate the next generation of optimized structural designs.

Over the years, numerous strategies have been developed to improve the computational efficiency of topology optimization (TO). Techniques such as reduced-order models \cite{amir2012efficient}, multi-grid conjugate gradient methods \cite{huang2017novel}, and parallel computing approaches \cite{aage2015topology} have all contributed to reducing the cost of solving large-scale TO problems. Yet, as the complexity of real-world designs continues to grow, these methods still face limitations in terms of scalability and the ability to handle multi-physics problems effectively. Recently, the intersection of machine learning (ML) and computational mechanics has brought new momentum to the field. Data-driven approaches have introduced novel ways to accelerate TO by leveraging predictive models \cite{shin2023topology, papadrakakis1998structural, banga20183d, abueidda2020topology, kallioras2020accelerated, li2019non}. These models, often neural networks (NNs), are trained on large datasets to map input parameters—such as loading conditions or boundary configurations—to optimal topologies \cite{ulu2016data}. This approach offers the potential for real-time prediction of optimized structures while significantly reducing the computational burden. Similarly, convolutional neural networks (CNNs) have been applied to predict optimal topologies from initial designs, generating near-instant results at a lower computational cost \cite{nie2021topologygan}. 
Sosnovik and Oseledets \cite{sosnovik2019neural} proposed a U-Net architecture to map early design iterations to final optimized structures, treating TO as an image segmentation task. Expanding on this, Yu et al. \cite{yu2019deep} introduced the idea of encoding boundary conditions as matrices, which improved the CNN's ability to generalize across different scenarios. Meanwhile, generative models such as conditional generative adversarial networks (CGANs) have been employed to refine coarse predictions into high-resolution topologies with more detailed features \cite{banga20183d}.
Despite these advances, data-driven methods face inherent limitations \cite{shin2023topology}. Large, labeled datasets are required for training, which can be time-consuming and computationally expensive to generate. Moreover, these models struggle to generalize to unseen boundary conditions or new loading configurations, leading to errors in predicted topologies \cite{zhou2020new}. Issues such as blurry edges, disconnected regions, and irregular shapes can reduce the effectiveness of the generated designs \cite{zhang2021tonr}. While online training methods \cite{deng2022self, chi2021universal, behzadi2021real} offer some improvements by using neural networks to represent the density field, they still rely on traditional finite element analysis (FEA) and struggle with distributed loads. 
These challenges pave the way for more robust approaches, such as physics-informed neural networks (PINNs) \cite{raissi2019physics}, which embed physical laws directly into the optimization process and offer a more scalable solution for complex TO problems \cite{rao2021physics}.

Recent approaches in topology optimization (TO) have sought to integrate these models as an alternative to traditional solvers \cite{zhang2021tonr, chen2021new, chandrasekhar2021tounn, ji2023recent, yin2024dynamically}. PINNs are particularly suited for TO because they reformulate the optimization process by embedding physical laws into the neural network's loss function. Instead of relying on conventional numerical methods to solve partial differential equations (PDEs), the problem is transformed into minimizing a residual or energy functional \cite{li2021physics, he2023deep}. This reparameterization of design variables as neural network parameters—such as weights and biases—allows for the use of deep learning backpropagation techniques, effectively replacing sensitivity analyses used in traditional TO methods \cite{zhang2021tonr}. By leveraging automatic differentiation, PINNs can compute gradients with a higher degree of precision than finite element analysis (FEA), which often suffers from numerical instability and floating point errors \cite{tijskens2002automatic,margossian2019review}.
One of the pioneering contributions in this area is the TOuNN approach proposed by Chandrasekhar and Suresh \cite{chandrasekhar2021tounn}, where the density function is parameterized by neural network parameters. The conventional FEA solver is embedded as an unconstrained loss function, where elemental stiffness and displacement are computed and used to calculate total compliance as part of the network’s training objective. Similarly, Chen and Shen \cite{chen2021new} introduced a physics-informed deep learning framework for density-based TO that updates the neural network using gradient information derived from FEA, rather than relying solely on data-driven training. This allows PINN-based TO methods to inherit the advantages of physics-informed learning, such as the ability to incorporate boundary conditions and governing equations directly into the model.
Despite these advancements, several challenges remain \cite{lu2021physics}. Many existing PINN-based TO methods \cite{jeong2023physics, jeong2023complete, chandrasekhar2021tounn, joglekar2023dmf, zehnder2021ntopo} are primarily limited to linear elastic structures, with only a few studies, such as Zhang et al. \cite{zhang2021tonr}, addressing geometric and material nonlinearities. Nonlinear TO problems often face numerical instabilities due to void elements with minimal stiffness, leading to indefinite stiffness matrices and computational difficulties \cite{kumar2021topology}. Additionally, while automatic differentiation provides accurate solutions to PDEs, the high computational cost of PINNs remains a significant limitation, especially for complex physical fields and dynamic loading conditions \cite{tijskens2002automatic}. These challenges suggest that while PINNs hold great promise, further improvements are needed to make them more computationally efficient and applicable to a broader range of engineering problems.

Inspired by the identified research gaps, we present a novel dual physics-informed neural network (PINN)-based topology optimization (TO) framework. The proposed approach solely relies on neural networks (NNs), eliminating the need for numerical solutions of sensitivity or stiffness matrices. A novel neural architecture for topology optimization is developed that involves a `displacement NN' for structural analyses and a `density NN' for topology optimization. These two networks are connected through a loss function defined by the variational energy formulation, with penalty terms for both total potential energy and volume fraction. The key features of the proposed approach includes,
\begin{itemize}
    \item \textbf{Decoupled Displacement and Density NNs for Efficiency}: By decoupling the displacement NN from the density NN, the framework simplifies the learning process and hence, is computationally efficient. These networks, connected via the loss function, allow for efficient simultaneous optimization without needing sensitivity analyses.
    \item \textbf{Enhanced Accuracy}: The displacement NN uses a sinusoidal representation network (SIREN) activation function, which is superior for computing higher-order derivatives compared to the hyperbolic tangent function typically used in PINNs \cite{sitzmann2020implicit, joglekar2023dmf}. This results in improved accuracy in structural analysis, especially for problems requiring more precise derivative calculations. Additionally, the use of Fourier feature mapping to define design variables enables the framework to handle complex design configurations, expanding its capability beyond simple topologies.
    \item \textbf{Simultaneous Training with a Relaxed Nested Approach for Speed}: Instead of the traditional nested approach—where the displacement NN is trained within the density NN training loop, leading to higher computational costs—this framework trains both NNs concurrently. Periodic updates to the density NN provide a computational advantage, reducing the overall training time while still delivering optimal results.
\end{itemize}
Overall, the proposed dual PINN-based framework opens new pathways for solving complex TO problems by eliminating the dependence on labeled data or traditional solvers (e.g., FEA) for sensitivity analysis. Case studies presented validate the framework’s effectiveness and efficiency across various TO problems, proving it as a viable alternative to existing data-driven and physics-driven TO methods.

The structure of the remaining paper is as follows: Section \ref{sec:TO_formulation} explains the SIMP based variational formulation for TO. Section \ref{sec:formulation} elaborates on the methodology and the architecture of the density and displacement neural networks, along with their respective components, with a particular focus on the integration of PINNs to capture optimal topology. Furthermore, the section includes the algorithm used in the study. Section \ref{sec:numerical} presents the validation and testing of the proposed approach through various numerical experiments. Finally, Section \ref{sec:conclusion} concludes the paper, discusses our findings, and suggests future research directions.

\section{Topology optimization formulation}
\label{sec:TO_formulation}

The objective of topology optimization (TO) is to identify the best material distribution within a prescribed design domain, accounting for boundary conditions and constraints, in order to achieve maximum stiffness or minimum compliance. Mathematically, TO problem can be posed as \cite{sigmund200199}
\begin{equation}
\left.\begin{array}{rl}
\underset{\rho}{\min} : & \mathcal{J}(\rho, \boldsymbol{u}(\rho))= \boldsymbol{u}^T \textbf{K}(\rho) \boldsymbol{u} \\
\text{subject to :} & \textbf{K}(\rho) \boldsymbol{u} = \boldsymbol{f}\\
: & \mathcal{C}(\rho) := \displaystyle \frac{V(\rho)}{V_0} - \mathcal{V}_f \leq 0,\\
: & 0<\rho_{\min } \leq \rho(\boldsymbol{x}) \leq 1, \; \boldsymbol{x}\in \Omega
\end{array}\right\}
\label{eq: optimization}
\end{equation}  
where, $\mathcal{J}$ represents the structural compliance, $\boldsymbol{u}$ represents the displacement field, $\rho$ represents the design parameters. \textbf{K} is the structural stiffness matrix, $\mathcal{C}$ represents the volume constraint imposed on the optimization process, $\mathcal{V}_f$ is target volume fraction, $\boldsymbol{x}$ is the coordinates of a point in design domain $\Omega$, and $V(\rho)$ and $V_0$ are the material and design space volume, respectively. Conventional TO process mainly comprises three steps: finite element analysis (FEA) of partial differential equations, sensitivity analysis, and design variable updating. The solid isotropic material with penalization (SIMP) method has been widely accepted as an effective tool for addressing topology optimization problems.
This method involves interpolating material properties within the design domain between a solid phase with maximum stiffness and a void phase with minimal stiffness. The interpolation is controlled by a penalized density variable, which encourages the creation of distinct solid and void regions. In the SIMP model, the Young’s modulus \( E(\rho) \) of each element is calculated based on its density \( \rho \) (where \( 0 \leq \rho \leq 1 \)). The Young’s modulus varies between a minimum value \( E_{\text{min}} \), representing voids, and a maximum value \( E_{\text{max}} \), representing solid material. The relationship is given by \cite{bendsoe1999material}
\begin{equation}
E(\rho) = E_{\text{min}} + \rho^p (E_{\text{max}} - E_{\text{min}})
\end{equation}
Here, \( E_{\text{min}} \) is the minimum Young's modulus associated with void regions, and \( E_{\text{max}} \) is the Young's modulus of the solid material. The penalization factor, $p$ plays a key role in mitigating intermediate density values, which helps to obtain a clear separation between solid and void regions in the final design. The penalization factor is commonly selected as $p=3$ in the SIMP method. In this study, we have also adopted $p=3$ for consistency with established practices. This interpolation scheme helps the optimization process find optimal material layouts that meet the required performance criteria while ensuring structural stability. By adjusting the penalization factor \( p \), designers can control how sharply the material transitions between solid and void, leading to desired structural designs. However, SIMP and other TO optimization algorithms are computationally expensive because of the presence of the nested loop. The objective of this paper is to propose a novel TO algorithm rooted in recent developments in the field of deep learning.

\section{Proposed approach}
\label{sec:formulation}
In this section, we present the core components of the proposed DPNN-TO framework and detail the training procedure employed. However, before proceeding with the details on the proposed neural architecture, we first illustrate how TO can be viewed as a energy minimization problem. 
At the end of the section, we provide an algorithm outlining the steps involved in implementing the method. Furthermore, the architectures of both the density and displacement neural networks are described comprehensively.
\subsection{TO as energy minimization}
\label{sec:governing_equation}
Our approach is based entirely on solving the energy minimization problem under specified constraints and boundary conditions. To that end, we formulate a variational integral functional expressed in terms of potential energy to accurately capture structural compliance. Let us consider an arbitrarily shaped linear elastic body $\Omega_{mat}$, which is a subset of a larger reference domain $\Omega \subset \mathbb{R}^d$, where $d = 2$ or $3$, and has boundary $\Gamma$. The displacement at a point $x \in \Omega$ is denoted by $\boldsymbol{u}(x)$. The boundaries of the domain $\Omega$ are classified as Dirichlet and Neumann boundaries, represented by $\Gamma_D$ and $\Gamma_N$, respectively.  
In the absence of inertial and body forces, the equilibrium equation can be expressed in tensorial form as
\begin{equation}
    \nabla\cdot \boldsymbol{\sigma} (\boldsymbol{x}) = 0, \;\; \forall  \boldsymbol{x} \in \Omega 
\end{equation}
where $\boldsymbol{\sigma}$ is the Cauchy stress tensor, and the system is subject to Dirichlet boundary conditions
\begin{equation}
    \boldsymbol{u} = \boldsymbol{u}_0, \;\; \forall  \boldsymbol{x} \in \Gamma_D
\end{equation}
where $\boldsymbol{u}_0$ is the prescribed displacement, and Neumann boundary conditions are expressed as
\begin{equation}
    \boldsymbol{\sigma} \cdot \boldsymbol{n} = \boldsymbol{f}, \;\; \forall  \boldsymbol{x} \in \Gamma_N
\end{equation}
where $\boldsymbol{n}$ is the outward unit normal to the boundary.
The total potential energy of the domain, $\Pi(\boldsymbol{u})$, depends on $u$ and can be computed as
\begin{equation}
    \Pi(\boldsymbol{u},\rho) = \mathcal{U}+\mathcal{V}
\end{equation}
where $\mathcal{U}$ represents the internal (elastic strain) energy, and $\mathcal{V}$ denotes the external (work potential) energy due to external forces. Considering isotropic linear elastic solid deformed under small strain conditions, the strain energy density function $\Psi(\boldsymbol{\epsilon({u})},\rho)$ is defined as
\begin{equation}
    \Psi(\boldsymbol{\epsilon({u})},\rho) = \frac{\lambda}{2} (tr [\boldsymbol{\epsilon({u})}])^2 + \mu tr[\boldsymbol{\epsilon({u})}^2]
\end{equation}
where, $\boldsymbol{\epsilon({u})}$ is the Cauchy strain which is function of $\boldsymbol{u}$, and $\lambda$ and $\mu$ are Lame's constants. For a plane stress problem, Lame's parameters can be calculated as
\begin{equation}
\lambda = \frac{E(\rho)\nu}{(1+\nu)(1-2\nu)} \; \; \; \text{and} \; \; \; \mu = \frac{E(\rho)}{2(1+\nu)}
\end{equation}
where $\nu$ is Poisson’s ratio and $E(\rho)$ is the modulus of elasticity which is the function of the design variable $(\rho)$. Using these definitions, internal energy (total strain energy) is defined as
\begin{equation}
    \mathcal{U}  = \int_\Omega \Psi(\boldsymbol{\epsilon({u})},\rho) \; \text{d}\Omega
\end{equation}
The external energy (work potential) is defined as the potential of the externally applied loads, which is the negative of the work done on the system. The work potential is given by
\begin{equation}
    \mathcal{V} = -\int_\Omega \boldsymbol{u}\cdot B d\Omega - \int_{\Gamma_N} \boldsymbol{u} \cdot f d\Gamma
\end{equation}
  
Since the effect of body forces ($B$) is neglected in our study, the external potential energy due to external forces can be expressed as
\begin{equation}
    \mathcal{V} = -\int_{\Gamma_N} u \cdot f d\Gamma 
\end{equation}
where $f$ is the external force (traction) vector applied on the boundary, $\Gamma_N$. Considering the equivalence between compliance and strain energy
\begin{equation}
    \int_\Omega \Psi(\boldsymbol{\epsilon({u})},\rho) \; \text{d}\Omega = \frac{1}{2} \int_{\Gamma_N} u \cdot f d\Gamma 
\end{equation}
Therefore, the total potential energy becomes:
\begin{equation}
    \Pi(\boldsymbol{u},\rho) = \int_\Omega \Psi(\boldsymbol{\epsilon({u})},\rho) \; \text{d}\Omega -  \int_{\Gamma_N} u \cdot f d\Gamma =  - \frac{1}{2} \int_{\Gamma_N} u \cdot f d\Gamma 
\end{equation}
Since compliance ($\mathcal{J}$) is defined as  $\int_{\Gamma_N} u \cdot f d\Gamma$ , we can express the compliance in terms of total potential energy
\begin{equation}
    \mathcal{J} = -2\Pi
\end{equation}
Therefore, the topology optimization problem can be reformulated in terms of the total potential energy as follows
\begin{equation}
\left.\begin{array}{rl}
\underset{\rho, \boldsymbol{u}}{\min} : & \mathcal{J}(\rho, \boldsymbol{u}(\rho))= - 2\Pi \\
\text{subject to :} & \mathcal{C}(\rho) := \displaystyle \frac{V(\rho)}{V_0} - \mathcal{V}_f \leq 0,\\
: & 0<\rho_{\min } \leq \rho(\boldsymbol{x}) \leq 1, \; \boldsymbol{x}\in \Omega
\end{array}\right\}
\label{eq: optimization1}
\end{equation} 
In summary, an objective function is established for the TO problem. Within the proposed framework, we will exploit the energy formulation to solve the TO problem by using physics-informed neural networks. 

\subsection{Dual PINN framework}

\begin{figure}
    \centering
    \includegraphics{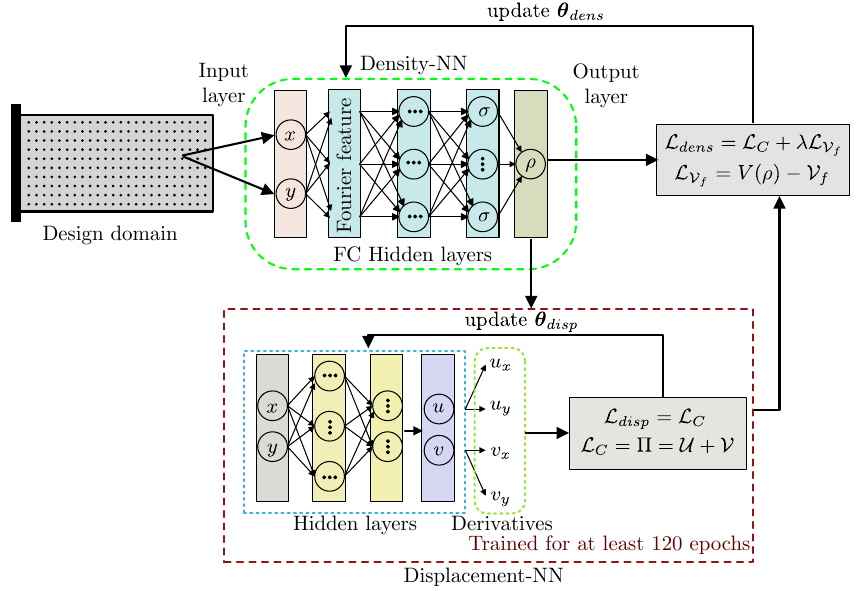}
    \caption{A dual physics-informed neural network-based framework for topology optimization (DPNN-TO), illustrating a two-dimensional computational domain having collocation points. These points represent the spatial coordinates used to train the displacement and density neural networks.}
    \label{fig:DPNN-TO}
\end{figure}
In the context of physics-informed neural networks (PINNs) applied to structural analysis, two primary methodologies are commonly utilized to solve partial differential equations (PDEs) \cite{samaniego2020energy, goswami2020adaptive, wang2023dcem}. The first approach minimizes the residuals of the PDEs and boundary conditions (BCs), ensuring that the neural network solution satisfies the governing equations. The second approach, known as the energy-based method, aims to minimize the system’s total potential energy \cite{goswami2020transfer}. Given its higher accuracy in the analysis of elastic structural systems, we adopt the energy-based method in this work by exploiting the formulation presented in Section \ref{sec:governing_equation}. 

This section outlines the key features of the DPNN-TO framework for structural topology optimization. Our approach integrates two neural networks, each contributing to different aspects of the optimization process. The first NN is tasked with calculating the displacement field of the structure, while the second NN estimates the material density distribution. Both networks accept spatial coordinates as input and output the corresponding displacement and density fields, respectively. The displacement and density fields are parameterized using physics-informed NNs, and these networks are coupled through a shared loss function, ensuring that the solutions to both fields are consistent with the underlying physics of the problem. The integrated architecture of the framework is illustrated in Fig.~\ref{fig:DPNN-TO}. 

\subsubsection{Displacement network design}
\label{sec:disp_NN}   
This subsection provides an overview of the displacement NN employed for finite element analysis. The displacement NN is designed to approximate the displacement fields resulting from applied loads, ensuring adherence to governing equations and boundary conditions. In this study, the focus is on linear elastic materials subjected to small deformations.

\begin{figure}
    \centering
    \includegraphics{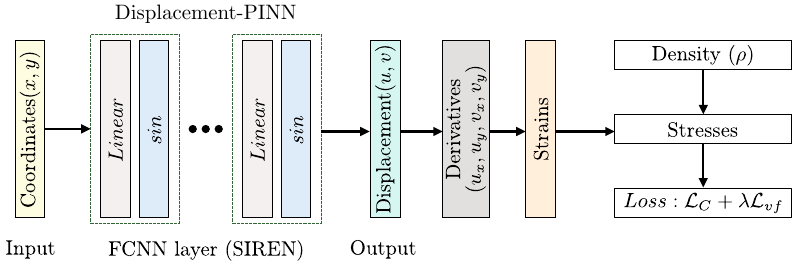}
    \caption{Architecture of displacement-NN. The neural network receives the spatial coordinates $x$ and $y$ as input. These coordinates are processed through multiple hidden layers incorporating linear transformations and a sinusoidal activation function, ultimately producing the corresponding displacement outputs $u$ and $v$. The loss term is calculated using the derivatives of displacements via automatic differentiation.}
    \label{fig:Disp-PINN}
\end{figure}

\textbf{NN architecture: }The architecture of the displacement network is structured to effectively model the complex relationships between input spatial coordinates and the resulting displacement fields. As illustrated in Fig.~\ref{fig:Disp-PINN}, the network consists of a series of fully connected layers (FCNN), including input, hidden, and output layers. The goal of the network is to map the spatial coordinates ($x,y$) to their corresponding displacement components ($u,v$). The output from the displacement NN can be formulated as 
\begin{equation}
    u, v = \mathcal{N}_{disp}(x, y; \boldsymbol{\theta}_{disp}),
\end{equation}
where $x$, $y$ represent the spatial coordinates and $\boldsymbol{\theta_{disp}}$ are learnable parameters of the displacement NN. The outputs $\boldsymbol{u}$ and $\boldsymbol{v}$ represent the component of the displacement field at the input coordinates. The forward propagation of the spatial coordinate data through the network is described layer by layer as follows: 
\begin{equation}
    h_j = \alpha(\mathbf W_j h_{j-1} + \bm b_j), \;\;\; j = 1, \cdots , N,
\end{equation}
where $N$ represents the total number of layers in fully connected network. $h_0$ is the input layer (representing spatial coordinates), and $h_N$ is the output layer (representing the displacement components). $\mathbf W_i$ and $\bm b_i$ are the weights and bias of the $i^{th}$ layer, respectively. $\alpha$ is the activation function that introduces nonlinearity, enabling the network to model a more complex relationship between inputs and outputs. 

Instead of the commonly used hyperbolic tangent (tanh) or rectified linear unit (ReLU) activation functions, we employ the sinusoidal representation network (SIREN) as the activation function,
\begin{equation}
    sin (h) = \frac{e^h-e^{-h}}{e^h+e^{-h}}
\end{equation}
This function provides the network with the ability to capture high-frequency features that are critical for applications involving derivatives, such as strain and stress calculations. This choice significantly improves the network's ability to compute derivatives of the output, which is essential for accurately determining strain and stress fields. Through this carefully constructed architecture, the displacement network effectively provides accurate displacement fields that satisfy the equilibrium conditions and boundary constraints imposed by the problem. This architecture serves as a key component in the proposed PINN-based topology optimization framework, facilitating precise and computationally efficient structural analysis.

\textbf{Loss function formulation: }
The loss function in the displacement NN is designed to ensure that the network output adheres to the principle of minimum potential energy. This principle states that the true displacement field of an elastic body minimizes the total potential energy of the system, aligning with structural mechanics principles. The displacement NN accomplishes this by minimizing the compliance, which is expressed in terms of the system’s internal strain energy. The loss function for displacement NN can be stated as:
\begin{equation}
    \mathcal{L}_{disp} = \mathcal{L}_C
\end{equation}
where $\mathcal{L}_C$ represents the normalized compliance loss, which is given by:
\begin{equation}
    \mathcal{L}_C = \frac{\mathcal{J}}{\mathcal{J}_0}
\end{equation}
where $\mathcal{J}$ represents the structural compliance of the current epoch, and $\mathcal{J}_0$ denotes the compliance value at the initial epoch. Normalizing compliance serves to regularize the loss term, ensuring consistency in the optimization process and appropriately scaling the compliance values relative to their initial state.
To solve the linear elastic problem, the output from the displacement NN is found by solving the following optimization problem:
\begin{equation}
\begin{array}{rl}
    u,v  & = argmin_{\boldsymbol{\theta_{disp}}}\mathcal{N}(x,y;\boldsymbol{\theta_{disp}}) \\
    \\
     \text {subject to: }  \boldsymbol{u}(x) &= \boldsymbol{u}_0, \;\; \forall  x \in \Gamma_D
\end{array}
\end{equation}
We adjust the output of NN to ensure that the Dirichlet BCs are automatically satisfied. To achieve this, we define,
\begin{equation}
    \boldsymbol{u} = \boldsymbol{u}_0 + \mathcal{B}(x,y) \; \mathcal{N} (x,y;\boldsymbol{\theta_{disp}})
\end{equation}
where $\mathcal{B}(x,y)$ is a function that vanishes on the Dirichlet boundary $\Gamma_D$ but takes positive values in the interior of the domain $\Omega$. This formulation ensures that the predicted displacement field satisfies the prescribed boundary conditions. 

\subsubsection{Density network design}
In this subsection, we present an overview of the density neural network, referred to as the neural reparameterization method.  The core objective of the density network is to effectively approximate the material density field while satisfying a predefined volume fraction constraint. This is achieved through the optimization of NN parameters. The density NN employs the Adam optimizer to minimize the loss function. By utilizing an NN framework, the density field is not only approximated but also continuously adjusted throughout the training process. The loss function encompasses total energy minimization, ensuring that the network outputs reflect the required density distribution and adhere to volume constraints.

\begin{figure}
    \centering
    \includegraphics{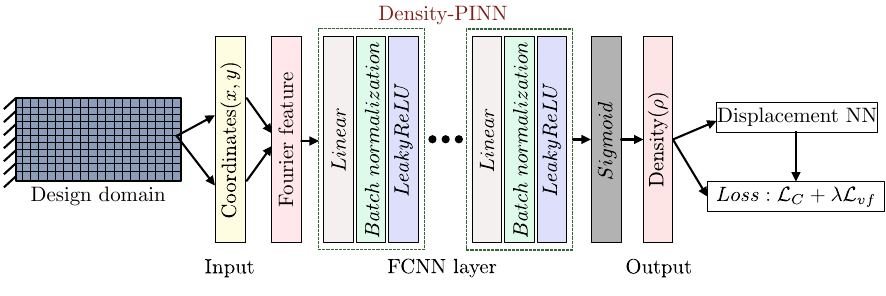}
    \caption{Architecture of the density-NN. The input consists of spatial coordinates $x, y$, which are first processed through a Fourier projection layer. The data then passes through a series of hidden layers, including linear transformations, batch normalization, and Leaky ReLU activation functions. The network produces the corresponding density field ($\rho$) as the final output.}
    \label{fig:Dens-PINN}
\end{figure}

\textbf{NN architecture: }
This network is designed to model the design variable $\rho$, as illustrated in Fig.~\ref{fig:Dens-PINN}. The formulation of the density NN can be expressed as follows:
\begin{equation}
    \rho = \mathcal{N}_{dens}(x, y; \boldsymbol{\theta}_{dens})
\end{equation}
where $\boldsymbol{\theta_{dens}}$ denotes the parameters associated with the density network. 
The network comprises several components, including input, hidden, and output linear layers, integrated with batch normalization layers, a Fourier projection layer, and a leaky rectified linear unit (LeakyReLU) activation function. This configuration enhances training stability and accelerates convergence speed. The LeakyReLU activation function is defined as follows:
\begin{equation}
    g (h) = max(0.01 h, h)
\end{equation}
where $h$ is the input to the function. This formulation indicates that the output is the greater of $0.01h$ or $h$, which provides a small, non-zero output for negative input values, thus helping to avoid the issues associated with ``dying" neurons in traditional ReLU functions. The final layer of the density NN employs a sigmoid activation function to constrain the output values of the design variable within the range of $0$ to $1$. This scaling ensures that the approximated density values are physically meaningful and effectively represent the material distribution. The sigmoid function is defined as:
\begin{equation}
    k (h) = \frac{1}{1+e^{-h}}
\end{equation}
As depicted in Fig.~\ref{fig:Dens-PINN}, the output $\rho$ is generated through forward propagation of NN. In this approach, the direct update of design variables in TO is substituted by adjusting NN’s parameters. Therefore, the output of density-NN is governed by NN parameters denoted as $\rho$, which are employed to define the optimization problem as presented in \ref{sec:TO_formulation}. It is essential to highlight that representing output $\rho$ as a continuous function of the spatial coordinates $\boldsymbol{x}$ and $\boldsymbol{y}$ enables the creation of a structure with smooth boundaries by incorporating high-resolution sample points into a trained density-NN model.

\textbf{Fourier projection layer: }
The Fourier projection layer is a critical component of the density NN architecture. Fourier projection helps to identify the higher frequency geometry in the low dimensional input space \cite{white2018toplogical}. It allows the network to project the material distribution onto a basis of Fourier modes, enabling the representation of complex patterns while maintaining control over the smoothness of the distribution \cite{tancik2020fourier}. This layer helps in reducing the occurrence of numerical instabilities and checkerboarding effects, which are common in traditional topology optimization methods \cite{doosti2021topology}. The two-dimensional Euclidean input is mapped to a $2 \times n_f$ dimensional Fourier space, where $n_f$ represents the selected number of frequencies. The length scale controls determine the range of frequencies:
\begin{equation}
    f \in \left [\frac{a}{l_{min}}, \frac{a}{l_{max}} \right]
\end{equation} 
where $a$ is the mesh edge length, $l_{min}$ and $l_{max}$ are the minimum and maximum length scales. The Fourier feature mapping enables NNs consisting of multiple hidden layers to express high-frequency geometries in low-dimensional domains, and it is represented as:
\begin{equation}
    \mathcal{F} = [cos(2 \pi \boldsymbol{B}^T \boldsymbol{x}) \; \; sin(2 \pi \boldsymbol{B}^T \boldsymbol{x})]^T
\end{equation}
where the component of the matrix $\boldsymbol{B} \in \mathbb{R}^{d \times m}$ is the Fourier basis frequency sampled from a normal distribution with $N(0, \sigma^2)$. The dimension $m$ is the size of the Fourier feature mapping. It is noted that standard deviation $\sigma$ enables the control of the detail of the network output. As sigma increases, it produces higher-frequency geometry, resulting in more intricate shapes. On the other hand, a smaller $\sigma$ yields simpler geometry. For this reason, the Fourier feature mapping can be used in the design density network to control the length scale. To constrain the $\boldsymbol{\rho}$ to be between 0 and 1, the last layer uses the sigmoid function expressed as 
\begin{equation}
    \boldsymbol{\rho} = \frac{1}{1+exp(\boldsymbol{z_{\rho} w_{\rho} +b_{\rho}})}
\end{equation}

\textbf{Loss function formulation: }
The loss function for the density-NN is composed of two main components, addressing both the compliance and volume constraints of the design, as illustrated in Fig.~\ref{fig:Dens-PINN}. This formulation is expressed as:
\begin{equation}
    \mathcal{L}_{dens} = \mathcal{L}_C + \lambda \mathcal{L}_{v_{f}}
\end{equation}
where $\mathcal{L}_C$ represents the normalized compliance loss, and $\mathcal{L}_{v_{f}}$ accounts for the volume constraint. The second loss term, $\mathcal{L}_{v_f}$, is calculated using the $L2$ distance function as follows:
\begin{equation}
    \mathcal{L}_{v_f} = \sum_{i = 1}^n \left (\frac{V(\rho_i)}{V_i}-\mathcal{V}_f \right )^2
\end{equation}
Here, $\lambda$ serves as a weighting factor, balancing the contributions of the compliance and volume constraint terms. This factor is manually tuned to ensure optimal loss convergence and the desired performance of the density-NN. Unlike the conventional neural reparameterization method, which considers only the compliance of the design as a loss term, the density-NN approach in topology optimization incorporates both compliance and volume constraints. The gradient of the loss function is computed using automatic differentiation, enabling the updating of the NN’s parameters through backpropagation during training. Upon convergence of the loss function, the resulting density field represents the optimal structure, satisfying both the minimum compliance and the volume constraint. Notably, the filtering scheme commonly applied in density-based topology optimization has not been integrated explicitly within the density-NN framework, as the resulting designs do not exhibit the checkerboard patterns typically mitigated by such filtering. Additionally, blurred regions in the design are minimized, yielding solutions closer to binary (0-1) material distributions, which result in well-defined structural boundaries in the final design.

\subsection{Training strategy}
\label{sec:training}
The training and testing phases are conducted using collocation points, which are generated at equidistant intervals within the design domain. Numerical integration is performed using the Monte Carlo method. In the DPNN-TO framework, the displacement and density NNs are coupled in a nested configuration, where both networks depend on each other’s outputs. Specifically, the displacement network is embedded within the density network, and the density network’s loss function incorporates a compliance loss term that depends on the displacement network's output. 

Conversely, the displacement network uses the density network's output as part of its input. The optimization begins by initializing the density network and generating its initial output. Using this initial density, the displacement network is trained for several epochs to compute static equilibrium displacements. Once the displacement network is sufficiently trained, the compliance loss is calculated and integrated into the density network's loss function. Subsequently, the density network is updated using this loss, followed by retraining the displacement network for progressively fewer epochs with each density update. This iterative process continues, refining both networks until convergence is achieved. 

The training involves sampling domain coordinates and performing a forward pass through the density network to generate the current topology. Simultaneously, the displacement network computes the corresponding compliance, which is fed into the density network's loss function. The loss is backpropagated, and a gradient descent step is performed to update the density network’s parameters.

In each training epoch within the DPNN-TO framework, the following steps are performed: 1) The displacement network is trained using the current density field, along with collocation points and boundary conditions, to calculate static equilibrium displacements, 2) Domain coordinates are then sampled, and a forward pass is conducted through the density network to produce the current topology output, while the displacement network computes the corresponding compliance. This compliance is fed into the density network's loss function, 3) Finally, the density network loss is backpropagated, and a gradient descent step is performed to update the weights of the density network, driving the optimization process forward.

The displacement-NN aims to solve the equilibrium equation based on the principle of minimum potential energy. For a body in static equilibrium, with no external body forces, the system's potential energy is given by: 
\begin{equation}
    \Pi(\boldsymbol{u},\rho) = \int_\Omega \Psi(\boldsymbol{\epsilon({u})},\rho) \; \text{d}\Omega -  \int_{\Gamma_N} u \cdot f d\Gamma 
\end{equation}
The loss function ($\mathcal{L}$) in displacement-NN can be defined identically as the potential of the system \cite{samaniego2020energy, nguyen2020deep}:
\begin{equation}
    \mathcal{L}(\boldsymbol{u}, \rho) = \Pi(\boldsymbol{u},\rho)
\end{equation}
The solution to the elasticity problem, as provided by the displacement-NN model, can be expressed as:  
\begin{equation}
    \boldsymbol{u} = arg \min_u \mathcal{L}(\boldsymbol{u})
\end{equation}
In this work, the Adam optimizer is used to train the model. The learning rate is another parameter that the user can change to achieve optimal framework performance. This optimization process is repeated in an iterative manner until the design satisfies the following stopping criterion
\begin{equation}
    \frac{N_{gray}}{N_{total}} \leq \tau
\end{equation}
where $N_{gray}$ is the number of gray elements that have densities ranging between 0.05 and 0.95, $N_{total}$ is the total number of elements, and $\tau$ is the maximum allowable error. We have set the value of the maximum allowable error to 5 \%. 

\subsection{Proposed algorithm}
\label{sec:alogorithm}
To conclude the algorithm of the DPNN-TO framework, the workflow of this approach has been summarized through an explicit implementation process (Algorithm~\ref{alg:cap}).

\begin{algorithm}
\textbf{Input:} $\Omega, nelx, nely, \nu, \tau, E_{max}, E_{min}, \mathcal{V}_f, penal, \lambda, lr_{disp}, lr_{dens}, epoch_{init}, epoch_{disp}, epoch_{dens}$

\textbf{Result:} The optimized topology structure $\boldsymbol{\rho_{opt}}$ and the plot of the final design
\caption{A Dual PINN algorithm for topology optimization} \label{alg:cap}

\begin{algorithmic}[1]

\State Generate the collocation points in the design domain ($\Omega$) for training
\State Initialize optimizers: $optimizer_{dens}$, $optimizer_{disp}$
\State Initialize parameters $penal = 3$; $\lambda = 50$; $epoch = 0$;
\State Initialize $\mathcal{N}_{dens}(\boldsymbol{x},\boldsymbol{y}; \boldsymbol{\theta_{dens}})$ with $xavier$ initialization method and get $\boldsymbol{\rho_{init}}$
\Repeat 
\State $\boldsymbol{u}, \boldsymbol{v} \leftarrow \mathcal{N}_{disp}(\boldsymbol{x}, \boldsymbol{y}; \boldsymbol{\theta_{disp}})$: Apply boundary condition
\State Calculate compliance loss ($\mathcal{J}$) using $\boldsymbol{\rho_{init}}$
\State Update $\boldsymbol{\theta_{disp}}$ $\leftarrow$ $optmizer_{disp}$
\Until{ $epoch_{init}$ iterations }
\State $\mathcal{J}_0 \leftarrow \mathcal{J}$ : $\mathcal{L}_C = \frac{\mathcal{J}}{\mathcal{J}_0}$
\Repeat 
\State Reset gradients of the optimizers

\State Train $\mathcal{N}_{dens}$,  $\boldsymbol{\theta_{dens}} \leftarrow \underset{\boldsymbol{\theta_{dens}}}{\mathrm{argmin }} \; \mathcal{L}_C(\boldsymbol{\rho}, \boldsymbol{u}, \boldsymbol{v}) + \lambda \mathcal{L}_{v_f}$
\State Update $optimizer_{dens}$ and calculate $\boldsymbol{\rho} \leftarrow \mathcal{N}_{dens}(\boldsymbol{x},\boldsymbol{y}; \boldsymbol{\theta_{dens}})$ 
\State Clip gradient for $\mathcal{N}_{dens}$ to avoid exploding gradients
\Repeat
\State Train $\mathcal{N}_{disp}(\boldsymbol{x},\boldsymbol{y}; \boldsymbol{\theta_{disp}})$ : $\boldsymbol{\theta_{disp}} \leftarrow \underset{\boldsymbol{\theta_{disp}}}{\mathrm{argmin }} \; \mathcal{L}(\boldsymbol{\rho}, \boldsymbol{u}, \boldsymbol{v}) $
\Until{ $epoch_{disp}$ iterations }
\State epoch $\leftarrow$ epoch + 1 ;
\If {$error \leq \tau$} $break$;
\EndIf
\Until{ epoch $<$ $epoch_{dens}$}
\State Plot final design $\boldsymbol{\rho}_{\textbf{\text {opt }}}$
\end{algorithmic}
\end{algorithm}

\section{Numerical examples}
\label{sec:numerical}

In this section, we present a series of numerical experiments designed to evaluate the robustness and computational efficiency of the DPNN-TO framework. The examples include classical benchmark problems as well as more complex cases that test the framework's flexibility and efficiency. The proposed methodology uses the PyTorch library to create neural networks (NNs). All numerical simulations are performed on an Nvidia Quadro P2200 GPU with 5 GB of VRAM. The experiments are structured to investigate the following key aspects:

\begin{enumerate}
    \item \textbf{Validation and computational performance}:
    \begin{itemize}
        \item \textbf{Comparison with standard SIMP methods}: DPNN-TO is validated by comparing the topology designs obtained with those from standard two-dimensional designs using SIMP methods.
        \item \textbf{Trade-off analysis}: A comprehensive trade-off analysis compares the DPNN-TO framework with the conventional SIMP-based TO method.
        \item \textbf{Parametric study}: Examines the impact of mesh resolution and NN size on the framework's performance.
    \end{itemize}
    \item \textbf{Design with passive elements}: The capability of the DPNN-TO framework is demonstrated by solving two-dimensional TO problems that include passive elements within the design domain.
    \item \textbf{Multi-constraints design}: A design problem incorporating stress and volume constraints is addressed, showcasing the framework's ability to handle multiple constraints.
    \item \textbf{Multiple loading conditions}: The framework optimizes a topology subjected to multiple loading conditions, illustrating its robustness in complex loading scenarios.
    \item \textbf{Extension to three-dimensional design}: The DPNN-TO framework's capability is extended to address three-dimensional topology optimization problems, showcasing its applicability to more complex and intricate design challenges.
\end{enumerate}

\subsection{Validation and computational performance}
\label{sec:validation}

\begin{figure}
    \centering
    \includegraphics{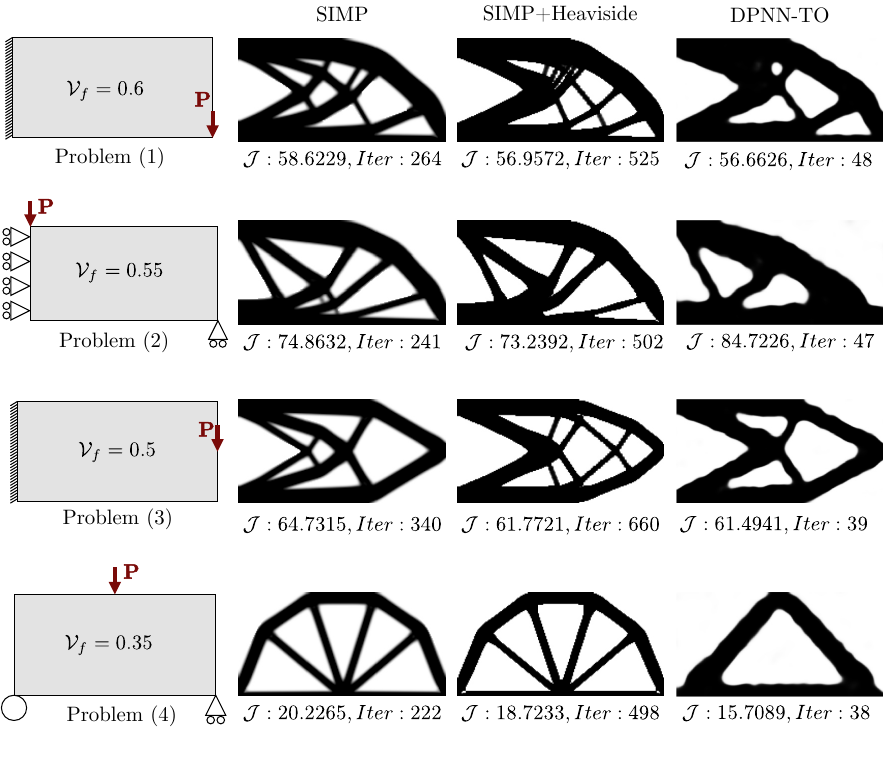}
    \caption{Optimized topologies for two-dimensional beam design problems. Presents the optimized topologies for four benchmark problems: (1) Cantilever beam with a load applied at the right bottom tip, (2) MBB beam, (3) Cantilever beam with a load applied at the right midpoint, and (4) Simply supported beam (SSB).}
    \label{fig:validation}
\end{figure}

In this section, we present a series of numerical experiments to assess the accuracy and efficiency of the proposed DPNN-TO framework and compare its performance against the SIMP-based methods. A collection of well-established benchmark problems in TO is used for this evaluation. The SIMP-based 88-line code \cite{andreassen2011efficient} serves as the reference for comparison. Initially, we focus on the classical two-dimensional compliance minimization problem, comparing the structural topologies produced by the DPNN-TO framework with those generated by the SIMP-based 88-line MATLAB implementation. The design domain for the two-dimensional beam is $2 \, \text{m} \times 1 \, \text{m}$ in size and is discretized into $200 \times 100$ collocation points. The material properties are defined with a Young's modulus $E = 1 \, \text{Pa}$, a Poisson’s ratio of $\nu = 0.3$, and a minimum Young's modulus $E_{\text{min}} = 1 \times 10^{-9} \, \text{Pa}$, to prevent from numerical instabilities. A vertical load of $1 \, \text{N}$ is applied to the  structure. For the TO, a penalization factor $p = 3$ is used, and the target volume fractions are provided in Fig.~\ref{fig:validation}.

The displacement NN is constructed with four fully connected layers, each comprising 25 neurons. The density NN employs a Fourier feature representation for the input layer, followed by two hidden layers, each consisting 25 neurons. The number of Fourier features ($n_f$) is set to 150, while the maximum number of training epochs for density NN is capped at 100. The number of epochs for the initial training of the displacement NN is set to 500, with 120 epochs allocated for further training after the initial stage. The learning rates are configured as $1 \times 10^{-4}$ for the displacement NN and $1 \times 10^{-2}$ for the density NN. A customized loss weight of $\lambda = 50$ is assigned to balance the scales of the two loss components in the density NN's loss function. 

For validation, we consider examples of a two-dimensional cantilever beam, a simply supported beam, and a Messerschmitt-Bölkow-Blohm (MBB) beam examples as depicted in Fig.~\ref{fig:validation}. We compare the results with those obtained using conventional SIMP-based methods to assess the efficacy of our proposed methodology.  Key performance metrics are summarized to highlight the advantages of DPNN-TO in terms of computational efficiency and optimization effectiveness. The findings indicate that the DPNN-TO framework exhibits superior performance and efficiency.

\begin{figure}
    \centering
    \includegraphics{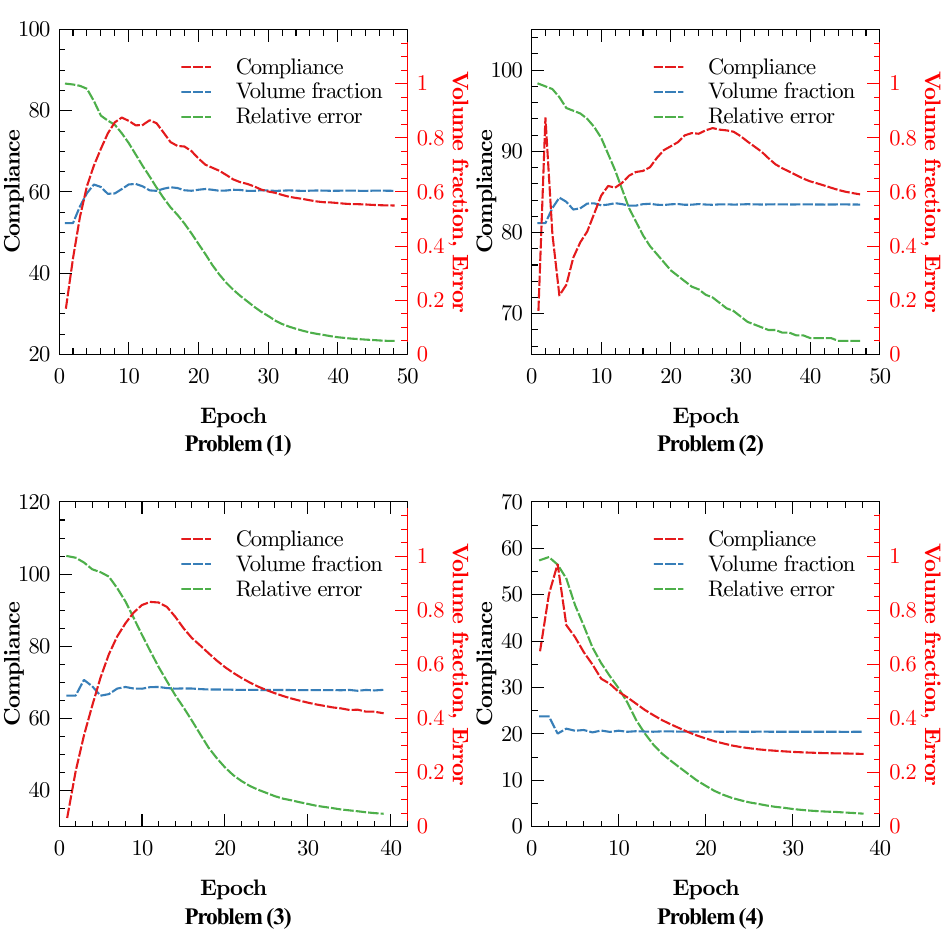}
    \caption{Convergence history plots of compliance, volume fraction constraint, and relative error.}
    \label{fig:loss1}
\end{figure}

The final topologies produced by DPNN-TO and SIMP show notable differences, as illustrated in Fig.~\ref{fig:validation}. Upon comparison, we found that our approach is superior as indicated by lower objective function values in Fig.~\ref{fig:validation}. SIMP tends to generate finer branch-like features, while DPNN-TO results in simpler structures with fewer voids. The topologies from DPNN-TO exhibit smooth and well-defined boundaries, in contrast to the blurred edges seen in SIMP designs. This blurring is due to the presence of gray elements, which represent intermediate density materials. These gray regions arise from overestimated strain energy in SIMP designs, leading to significant deviations in compliance compared to fully discrete structures. To facilitate a more direct comparison, the Heaviside projection filter is applied to SIMP designs, producing near-binary (0-1) solutions. It is also worth noting that DPNN-TO does not require filter schemes, as the output naturally achieves a near-discrete structure, free of the checkerboard pattern often observed in SIMP results. Additionally, high-resolution boundaries in DPNN-TO can be obtained by increasing the resolution of the sampling points in the trained model without incurring additional computational costs. 
By fine-tuning the hyperparameters of DPNN-TO, the model converges towards an optimal solution. The convergence and error plots for DPNN-TO are presented in Fig.~\ref{fig:loss1}. Initially, the compliance value exhibits a rise followed by a decline as the volume fraction approaches its target value. Notably, the volume fraction converges more rapidly than compliance, primarily due to the tailored loss weighting factor, $\lambda$, applied to the loss function of the density NN. This weighting results in the network parameters being more responsive to variations in the volume loss compared to the compliance loss.
The results for all four examples are summarized in Table~\ref{tab:comparison}, providing a quantitative basis for comparison between the methods.

\begin{figure}
    \centering
    \includegraphics{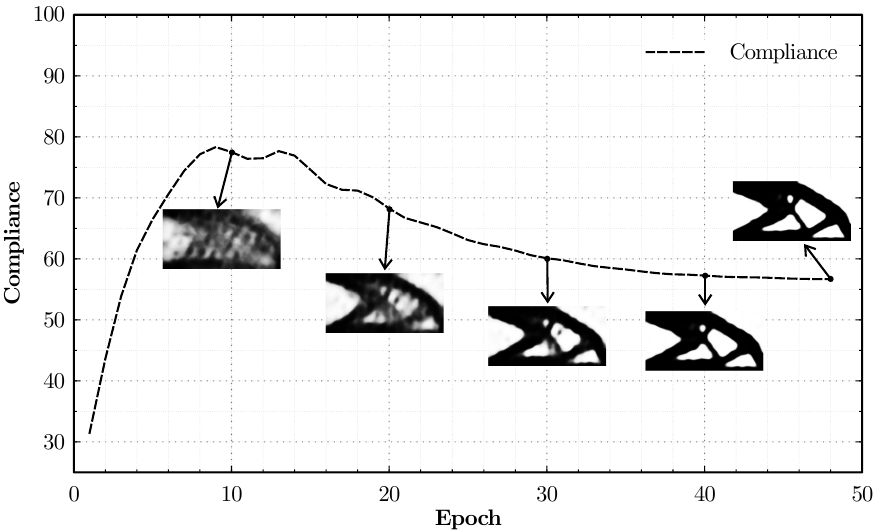}
    \caption{The convergence history of the problem (1) as shown in Fig.~\ref{fig:validation}.}
    \label{fig:convergence1}
\end{figure}

Furthermore, Fig.~\ref{fig:convergence1} presents the evolution of the optimal structural topology over successive epochs. The final optimized solution exhibits significant improvements in structural performance, as evidenced by the reduction in the objective function. This result underscores the effectiveness of our approach in delivering topologies that not only minimize compliance but also adhere to volume constraints.

\subsubsection{Computational performance and trade-off analysis}
\label{sec:trade-off}
In this section, we examine the computational efficiency of the proposed DPNN-TO framework and analyze the inherent trade-offs involved in achieving optimal topologies. We present a trade-off analysis focusing on key metrics such as objective function values (structural compliance) and the number of epochs required for convergence. Table ~\ref{tab:comparison} summarizes the comparison of these metrics between the DPNN-TO and the SIMP methods with Heaviside projection. The compliance values reflect the structural performance of each solution, while the epochs indicate the computational cost to reach convergence. Our analysis shows that DPNN-TO consistently produces topologies with lower compliance values, signifying superior structural performance. This improvement is especially pronounced in problems involving complex geometries.

\begin{table}
\centering
\caption{Comparison of the proposed method and the conventional SIMP method (88-line code), based on compliance value and iterations or epochs.}
\begin{tabular}{c|ccc|ccc}
\toprule
Case &  \multicolumn{3}{c|}{Compliance} & \multicolumn{3}{c}{Epochs}\\
\midrule
{}   & SIMP   & SIMP+Heaviside    & DPNN-TO   & SIMP   & SIMP+Heaviside    & DPNN-TO   \\
(1)   &  58.6229 & 56.9572 & 56.6626 & 264 & 525 & 48  \\
(2)   &  74.8632 & 73.2392 & 84.7226 & 241 & 502 & 47  \\
(3)   &  64.7315 & 61.7721 & 61.4941 & 340 & 660 & 39  \\
(4)   &  20.2265 & 18.7233 & 15.7089 & 222 & 498 & 38  \\
\bottomrule
\end{tabular}
\label{tab:comparison}
\end{table}

\begin{figure}
    \centering
    \includegraphics{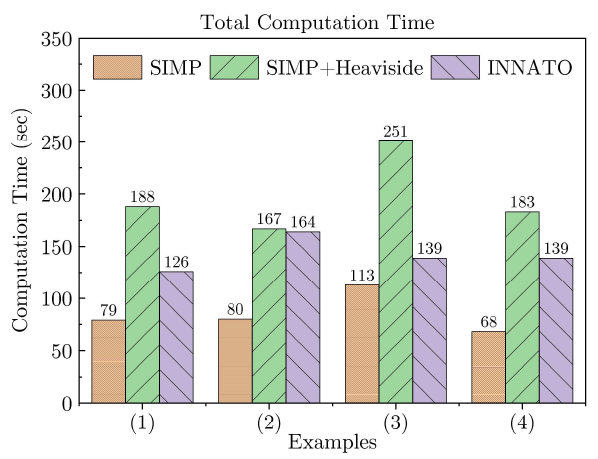}
    \caption{The total computational time in seconds for the problems presented in Fig.~\ref{fig:validation}.}
    \label{fig:computation_time}
\end{figure}

Fig.~\ref{fig:computation_time} shows the comparison of computational time, highlighting the performance differences between the approaches across different optimization cases. The compliance values demonstrate the effectiveness of each method in achieving optimal solutions, with DPNN-TO consistently achieving competitive results. The DPNN-TO framework demonstrates better computational efficiency compared to the SIMP with the Heaviside projection filter method, though it incurs higher costs when compared to the standard SIMP method. The high computational demand arises from the inclusion of two NNs, the displacement NN and the density NN. Contributing factors include the complexity of the network architectures, the number of training epochs, and the resolution of the design domain. For benchmark problems, especially those involving linear elasticity with small deformations, it is important to note that PINN generally takes more time to compute compared to conventional methods like FEA. In summary, this trade-off analysis provides valuable insights into the strengths and limitations of the DPNN-TO framework, offering a promising alternative for the TO process that demands efficiency and robustness in handling complex design constraints and loading conditions.

\subsubsection{Parametric Study}
\label{sec:parametric}
This section investigates the influence of various hyperparameters of DPNN-TO, with a particular focus on the effect of neural network size and the number of collocation points.

\begin{itemize}
    \item \textbf{NN Size dependency:} Estimating the optimal size of a neural network, in terms of depth and width, is a complex task due to the highly nonlinear nature of its parameter space. The size of NN, defined by the number of layers (depth) and the number of neurons per layer (width), critically impacts the accuracy and performance of the proposed DPNN-TO framework. Larger networks, with greater depth and width, are better equipped to capture complex, nonlinear features within the design domain. However, this increased capacity comes with trade-offs, such as higher computational costs, longer training times, and a greater risk of overfitting. Given that several studies have already explored the optimal size of the displacement NN, we focus on systematically investigating the impact of varying the size of the density NN. The results indicate that while increasing the network size typically enhances the accuracy of the predicted topologies, this improvement is not linear. For instance, as shown in Fig.~\ref{fig:para_NN_size}, when the network depth is increased from one to four layers, along with a corresponding increase in the number of neurons per layer, we observed a significant improvement in the quality of the optimized topology. However, further increases in network size resulted in diminishing returns, with only marginal improvements in accuracy and substantial increases in computational cost. Furthermore, careful hyperparameter tuning becomes more critical as the network size increases. We find that larger networks are more susceptible to overfitting and require precise adjustments of learning rates and regularization parameters to ensure generalization performance. The size and architecture of the neural networks used in DPNN-TO significantly affect their capability to approximate complex relationships within the design space.

    \begin{figure}
    \centering
    \includegraphics{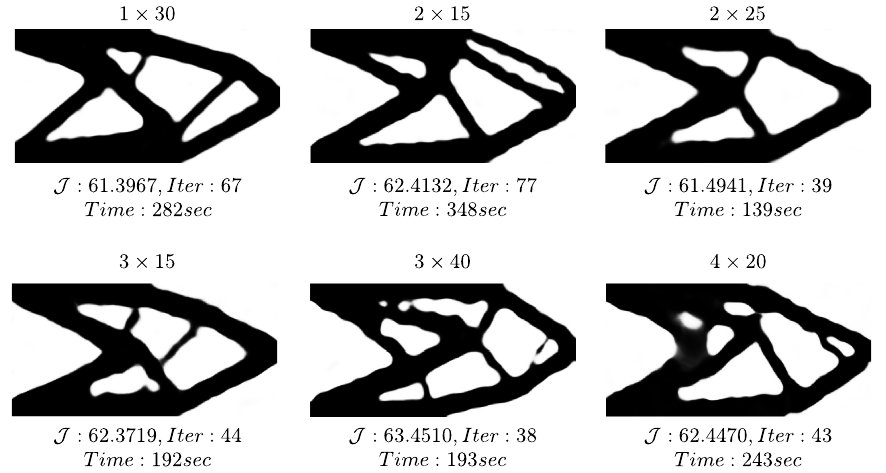}
    \caption{The optimal topologies obtained using DPNN-TO framework under various NN sizes (hidden layers $\times$ hidden nodes) for density NN.}
    \label{fig:para_NN_size}
    \end{figure}
    
    \item \textbf{Impact of mesh resolution:} Next, we examine the effects of mesh resolution (collocation points) on the performance of the DPNN-TO framework. A finer mesh, characterized by a higher number of collocation points, enables a more detailed and accurate representation of the design domain. This increased resolution improves the precision of material distribution, potentially resulting in more optimal topologies. However, finer meshes also come with higher computational demands, including longer convergence times and greater resource consumption.

    Conversely, a coarser mesh simplifies the domain representation and reduces computational costs. While this approach decreases the time and resources required for computation, it may lead to less accurate optimization outcomes, especially in regions that require detailed material distribution. Our study investigates how varying mesh resolutions impact the performance of DPNN-TO. We find that finer meshes generally yield more precise and refined topologies, enhancing the quality of the optimization results as shown in Fig.~\ref{fig:para_mesh_size}. However, this comes at the expense of increased computational cost and time. In contrast, coarser meshes alleviate the computational burden but may produce suboptimal results. After analyzing different resolutions, we settled on using $200 \times 100$ collocation points. 

    \begin{figure}
    \centering
    \includegraphics{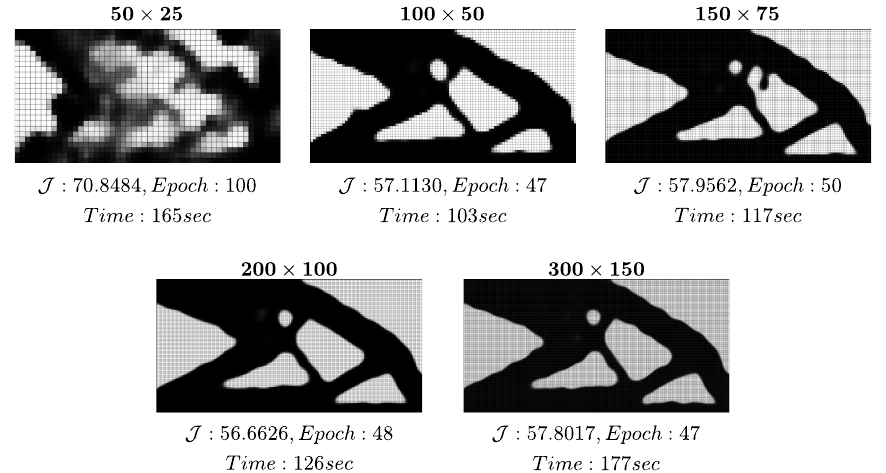}
    \caption{The optimal topologies obtained using the DPNN-TO framework under various mesh sizes.}
    \label{fig:para_mesh_size}
    \end{figure}

\end{itemize}

By systematically evaluating these parameters, this parametric study aims to improve the understanding and practical application of the DPNN-TO framework in engineering design. The findings contribute to optimizing the framework's efficiency and effectiveness, thereby facilitating its adoption for a broad range of TO challenges in engineering practice.

\subsection{Design with passive elements}
\label{sec:passive_element}
\begin{figure}
    \centering
    \includegraphics{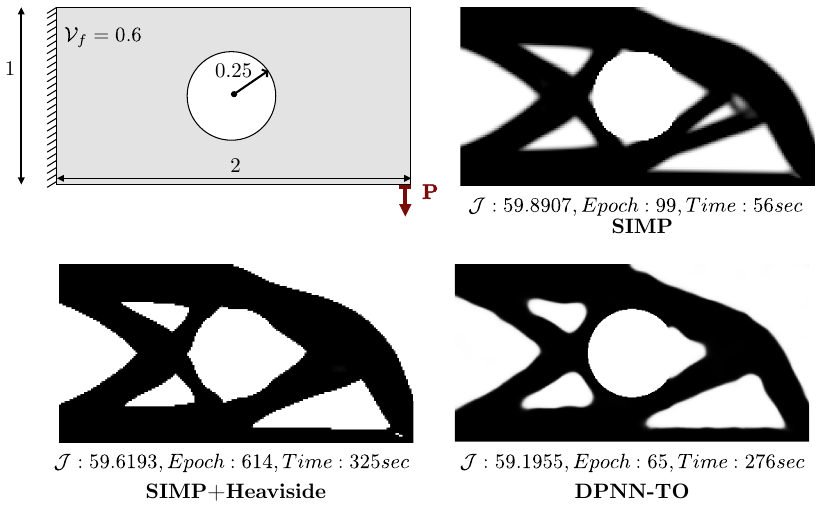}
    \caption{Topology optimization with passive elements using DPNN-TO framework. The cantilever beam design domain is subjected to an external load at the right-bottom tip. The circular region, with a diameter of 0.25, represents the passive element with fixed zero density (voids), demonstrating the integration of irregular geometries within the optimization framework.}
    \label{fig:passive}
\end{figure}

To further assess the capabilities of our DPNN-TO framework, we conducted TO in a design domain that incorporates passive elements. These passive elements, defined as regions with fixed densities of either 0 (void) or 1 (solid), were introduced to represent voids or solid regions within the computational domain during the optimization process, as illustrated in Fig.~\ref{fig:passive}. This was accomplished by assigning zero densities to the sample points within the specified void regions, effectively creating irregular geometries while maintaining consistent boundary conditions. In our methodology, passive elements are introduced by imposing zero density values at specific sample points within designated circular regions. This approach allows for the transformation of uniformly distributed sample points into irregular geometries while ensuring that the boundary conditions enforced by the PINNs remain consistent. Notably, the potential energy contribution of passive elements remains zero throughout the optimization process, ensuring minimal interference with the overall structural optimization objectives. 

Fig.~\ref{fig:passive} shows the optimized topology for a cantilever beam having a passive region inside it. We represented these regions as void and hence used zero densities during the optimization process. The dimension of the beam is 2 m $\times$ 1 m, and the diameter of the hole is 0.25 m. We marked the value of densities inside the circle as zero all the time. The mechanical properties and network hyperparameters used in this example are consistent with those discussed in the earlier validation case studies. The resulting optimal topology, as shown in Fig.~\ref{fig:passive}, demonstrates the ability of the DPNN-TO framework to to allocate material densities effectively outside the vacant regions, achieving the optimal structural layout solution. Incorporating passive elements into the design domain resulted in a 4.47\% increase in compliance compared to the design without passive elements. Despite this slight increase in compliance, the framework successfully accommodated the passive elements while optimizing the overall topology, showcasing its robustness and flexibility in handling complex design constraints.

The optimal topology is derived using the DPNN-TO framework, where void and solid densities are strategically allocated around the passive elements to optimize structural performance while adhering to design constraints. This approach not only enhances the flexibility of the TO process but also accommodates complex design requirements by seamlessly integrating predefined structural features. The inclusion of passive elements illustrates the capability of our methodology to handle diverse engineering challenges, from optimizing irregular geometries to incorporating functional constraints within the design domain.

\subsection{Multi-constraints design}
\label{sec:multiconst}

\begin{figure}
    \centering
    \includegraphics{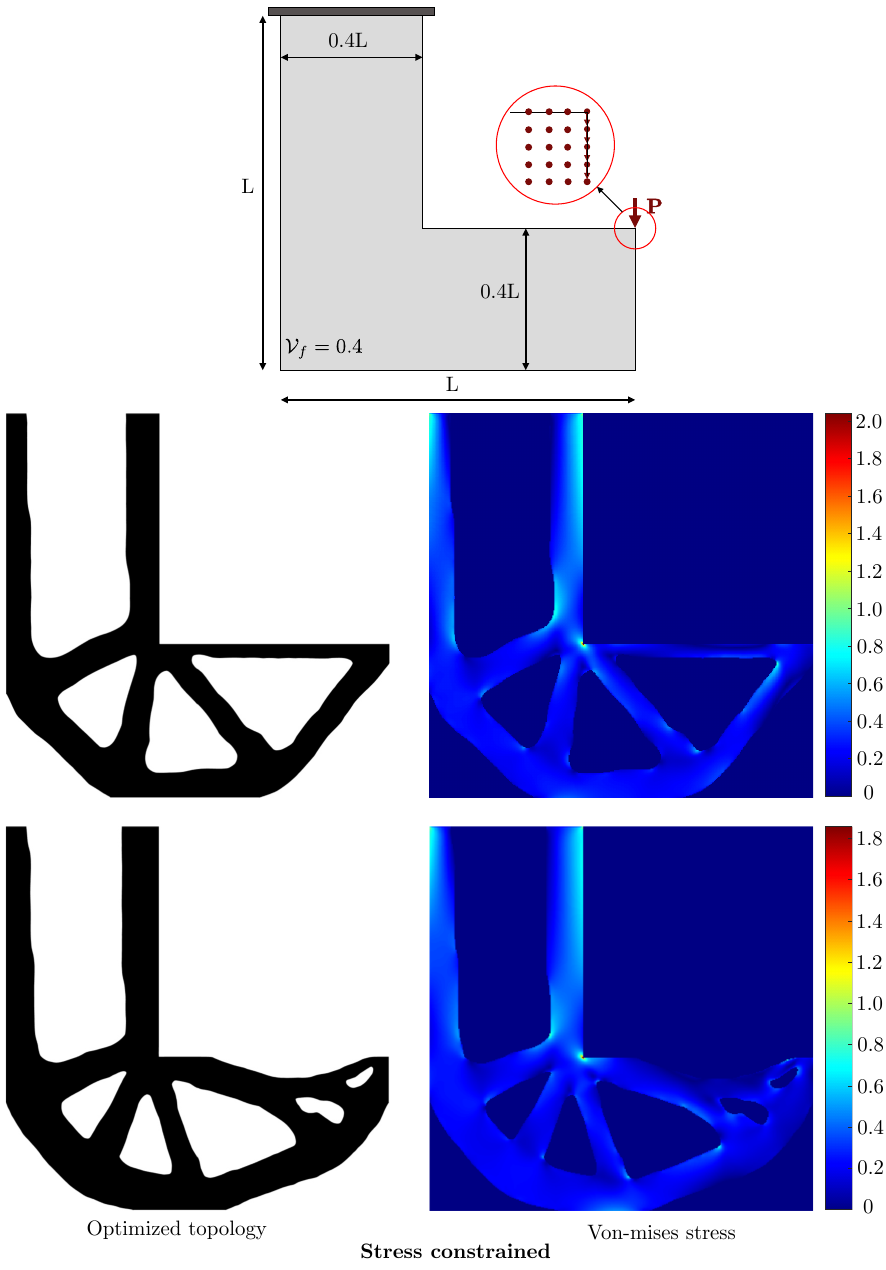}
    \caption{Topology optimization incorporating multiple constraints through the DPNN-TO framework. The design domain features an L-shaped beam subjected to an external load applied at the top right corner. This domain includes uniformly distributed sample points, illustrating the integration of both volume and stress constraints within the optimization process.}
    \label{fig:multicons}
\end{figure}

In this section, we extend our proposed DPNN-TO framework to tackle a compliance minimization problem that incorporates both volume fraction constraints and upper bounds on von Mises stress. The maximum von Mises stress is evaluated using the aggregated p-norm function. This p-norm stress measure provides a means to assess the maximum stress, where the condition $p_N \rightarrow \infty$ corresponds to the peak stress value; however, this condition may not hold physical relevance. The optimization problem for the associated stress constraint can be formulated as follows:

\begin{equation}
    \begin{split}
        min : \; & \mathcal{J}(\rho, \bd{u}(\rho))=-2\Pi;\\
        s.t. : \; &  \sigma_{vm}^{max} \leq \sigma_{vm}^*\\
        & V(\rho)/V_0 = V_f,
    \end{split}
\end{equation}

where $\sigma_{vm}^{max}$ denotes the maximum von Mises stress, and $\sigma_{vm}^{*}$ represents the specified value of stress constraint. To incorporate this stress constraint into the DPNN-TO framework, the L2 distance between the $\sigma_{vm}^{max}$ and $\sigma_{vm}^{*}$ denoted as $\mathcal{L}_{\sigma_{vms}}$ is included in the loss function. This approach enables the model to produce a structure that effectively minimizes the maximum von Mises stress in regions of stress concentration, adhering to the specified stress constraint. Specifically, this framework facilitates the generation of an optimized design with values approaching $0-1$, thereby mitigating the singularity issues that can arise in elements with low densities. As a result, the optimized structure adeptly addresses the problem of stress singularities within this framework. Given the highly non-linear characteristics of the stress-constraint topology optimization, a relatively small loss weight of 0.1 is assigned to the stress loss term. The multi-constraint optimization problem is evaluated using an L-bracket design subjected to an external load intensity of 1 $N/m^2$ that is distributed over 5 \% of the length of the beam. The material properties and topology optimization parameters are consistent with those discussed in the preceding sections. The target volume fraction $V_f$ of the L-bracket design is set to 0.4. In the p-norm function, a value of $p_N = 6$ is selected to effectively approximate the maximum von Mises stress. The maximum stress constraint, denoted as $\sigma_{vm}^*$, is set at $1.5 MPa$. The optimization loop terminates for the multi-constraint topology optimization process when the framework meets the following criteria: the maximum von Mises stress remains below the specified constraint value, and the design achieves the designated convergence error threshold. The network configuration for the displacement and density NN consists of $3 \times 64$ and $2 \times 64$ layers, respectively. The mesh size of the L-bracket is discretized into $200 \times 80$ points along the long and short edges. Due to a sharp corner in the design domain, a significant stress concentration is anticipated at this re-entrant corner. 

The compliance value is determined using the 88-line code to facilitate a comparative assessment of optimized topologies. The proposed DPNN-TO framework is adept at optimizing topology under stress constraints, yielding smoother boundaries that help to prevent stress singularities. As illustrated in Fig.~\ref{fig:multicons}, this framework effectively optimizes the structural topology of an L-shaped beam subjected to external loading while concurrently satisfying established volume and stress constraints. It can be observed that the maximum von Mises stress ($\sigma_{vm}^{max}$) has been reduced to 1.67 MPa from 2.04 MPa. Such capabilities enable the design of robust and efficient structures that fulfill multiple performance criteria.

\subsection{Multiple loading conditions}
\label{sec:multiload}
\begin{figure}
    \centering
    \includegraphics{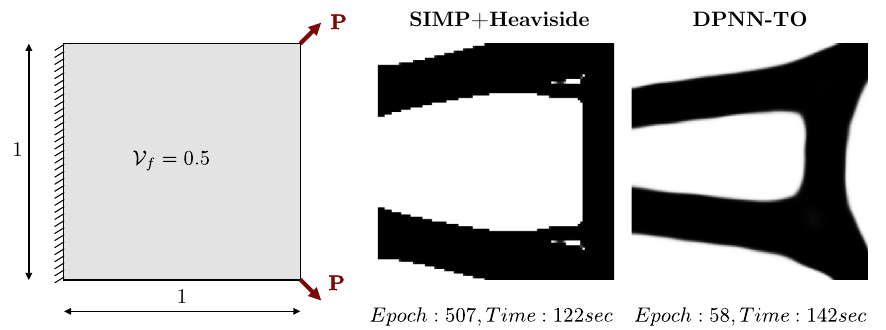}
    \caption{Topology optimization with multiple loads using the DPNN-TO framework. A short cantilever beam design domain is subjected to two inclined loads at the right bottom and top corners. The design domain comprises uniformly distributed sample points, demonstrating the integration of multiple loading conditions.}
    \label{fig:multiload}
\end{figure}

This section highlights the robustness of the DPNN-TO framework in optimizing structural designs under various loading conditions. To evaluate our approach further, we extend our methodology to solve a compliance minimization problem with multiple loads. In this case, two inclined loads are applied at the right corners of the cantilever beam design domain, as shown in Fig.~\ref{fig:multiload}. We have considered the collocation points $100 \times 100$ and the target volume fraction ($\mathcal{V}_f$) as 0.5 for this problem. The other parameters remain the same as in the previous validation cases.

Fig.~\ref{fig:multiload} shows the application of our DPNN-TO framework in optimizing the structural topology of a cantilever beam subjected to multiple external loads. The methodology effectively manages multiple loading scenarios by distributing material within the design domain to minimize compliance while maintaining structural integrity. By optimizing the distribution of material densities across the design domain, our approach enhances the structural performance of the cantilever beam, demonstrating its capability to address multi-loading challenges in engineering applications.

\subsection{Three-dimensional (3-D) example}
\label{sec:3D}
In this section, we extend the DPNN-TO framework to solve the three-dimensional optimization problem to check the scalability and robustness of the DPNN-TO framework. We optimize a three-dimensional cantilever beam, which introduces additional complexities, particularly in handling stresses and strains in the $z$-direction. This analysis aims to demonstrate the framework's capability to manage the challenges associated with three-dimensional structural optimization. Modifications have been made to the NN architecture to accommodate three-dimensional problems, including an additional output to approximate the displacement field in the $z$-direction. This ensures an accurate representation of structural deformations in all dimensions. Solving three-dimensional problems increases computational time due to the larger number of NN parameters and input data points involved. Nonetheless, fundamental hyperparameters such as learning rate, optimizer choice, and topology optimization parameters remain consistent with those used in the two-dimensional cases. 

\begin{figure}
    \centering
    \includegraphics{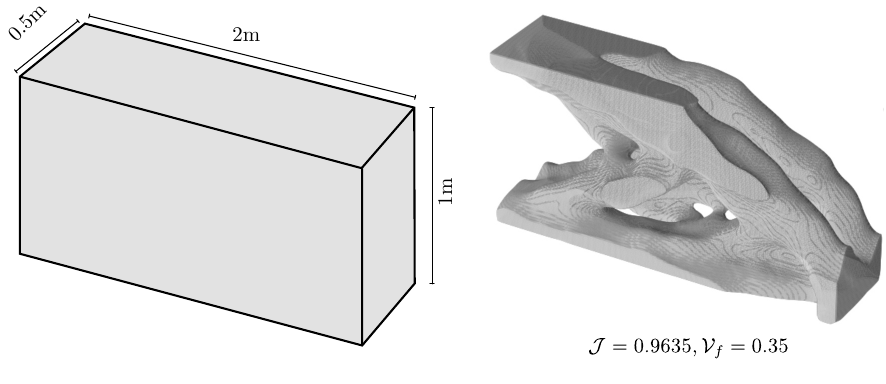}
    \caption{Topology optimization for a 3D cantilever beam using the PINN-based approach, loaded at the right bottom tip.}
    \label{fig:3D}
\end{figure}

To validate the three-dimensional implementation of the DPNN-TO framework, a cantilever beam problem subjected to distributed loading was considered, as depicted in Fig.~\ref{fig:3D}. In this example, the design domain was defined with dimensions of \(2 \, \text{m} \times 1 \, \text{m} \times 0.5 \, \text{m}\), and it was discretized into a grid of \(200 \times 100 \times 50\) points. This fine discretization was chosen to accurately capture the complex geometry and loading conditions of the problem, ensuring a high-resolution representation of the optimized topology. The optimization process for three-dimensional problems within the DPNN-TO framework follows a methodology similar to two-dimensional cases. The target volume fraction ($\mathcal{V}_f$) is set to 0.35. In this particular case, the number of Fourier features was set to 128 to enhance the expressive capacity of the density-NN. 

The NN architecture employed in this study was carefully designed to handle the increased complexity associated with three-dimensional problems. Specifically, the displacement-NN was configured with four layers, each containing 64 neurons, while the density-NN comprised three layers with 64 neurons each. This configuration was selected based on extensive hyperparameter tuning to balance computational efficiency with the accuracy of the optimization results. The model was trained using a batch size of 50000, which provided a good balance between computational time and convergence stability. The training process was conducted over a substantial number of epochs to ensure that both the compliance and volume fraction objectives were effectively minimized.

The optimized three-dimensional solution and the corresponding compliance values are illustrated in Fig.~\ref{fig:3D}. The resulting topology was visualized using the marching cubes algorithm, a well-suited technique for generating high-resolution isosurfaces from volumetric data. This approach allowed for an accurate and detailed representation of the optimized topology, capturing the intricate details of the material distribution within the design domain. The validation of the three-dimensional cantilever beam example demonstrates the robustness and versatility of the DPNN-TO framework in handling complex three-dimensional topology optimization problems. The ability to extend the framework from two to three-dimensional while maintaining computational efficiency and optimization accuracy is a significant achievement, highlighting the potential of DPNN-TO in practical engineering applications.

\section{Conclusion}
\label{sec:conclusion}
In this work, we have introduced the dual physics-informed neural network for topology optimization (DPNN-TO), a novel framework that combines two NNs: one for mapping the density field and the other for approximating the displacement field. We employ two neural networks with physics-informed loss functions as an alternative to the conventional solid isotropic material with penalization (SIMP) method for topology optimization. Instead of relying on finite element analysis, we use the displacement NN, which minimizes the total potential energy as its loss function. Minimizing the potential energy is equivalent to compliance minimization or maximizing structural stiffness. To update the density distribution, we use a dedicated density NN. This network is designed to ensure that the volume constraint is explicitly considered as part of its loss function. By doing so, the density NN helps to maintain the desired material distribution while adhering to volume constraints throughout the optimization process. DPNN-TO provides more realistic results for the different BC and loading conditions. We have validated our approach for TO across a diverse set of numerical examples. Our results demonstrate that the method consistently converges to the optimal solution, regardless of varying boundary conditions or loading scenarios. In particular, the output density field inherently achieves a smooth boundary structure, eliminating the occurrence of checkerboard patterns. This is accomplished by expressing the continuous density field as a function of input coordinates $x$ and $y$ through the density NN. Additionally, the DPNN-TO framework operates without the need for sensitivity filters or complex analytical formulations. The primary conclusions drawn from the study are elucidated as follows:
\begin{itemize}
    \item It can be observed that this framework produces simplified solutions without the formation of fine branches, which are typically prone to fracture. This contributes to the structural integrity and reliability of optimized designs.
\item The framework produces solutions with smooth boundaries, effectively avoiding the checkerboard patterns often encountered in traditional methods.
\end{itemize}

The numerical results of our study clearly demonstrate the effectiveness and robustness of the proposed DPNN-TO framework. 
This approach not only improves the accuracy of the optimization process but also provides a versatile framework capable of accommodating complex material behaviors. While this work marks a significant advancement, it also highlights existing challenges and presents opportunities for future research.  Although this framework can address complex optimization problems, it struggles to achieve optimal solutions in cases of stress-constraint topology optimization. The sensitivity of the framework to hyperparameters and the need for further validation with real-world experimental data are areas that earn attention.

Additionally, exploring the application of the DPNN-TO framework to different engineering domains and expanding the complexity of material models are avenues for future investigation. In conclusion, the proposed framework offers a promising approach to address the challenges posed by complexity in topology optimization. The integration of dual PINN with structural design opens new possibilities for the optimization of structures in diverse engineering applications.

\section*{Acknowledgment}
AS acknowledges the financial support received from the Ministry of Education, India, in the form of the Prime Minister's Research Fellows (PMRF) scholarship. SC acknowledges the financial support received from Anusandhan National Research Foundation (ANRF) via grant no. CRG/2023/007667 and from the Ministry of Port and Shipping via letter no. ST-14011/74/MT (356529).

\newpage
\appendix


\begin{thebibliography}{10}

\bibitem{bendsoe1988generating}
Martin~Philip Bends{\o}e and Noboru Kikuchi.
\newblock Generating optimal topologies in structural design using a homogenization method.
\newblock {\em Computer methods in applied mechanics and engineering}, 71(2):197--224, 1988.

\bibitem{bendsoe1999material}
Martin~P Bends{\o}e and Ole Sigmund.
\newblock Material interpolation schemes in topology optimization.
\newblock {\em Archive of applied mechanics}, 69:635--654, 1999.

\bibitem{sigmund200199}
Ole Sigmund.
\newblock A 99 line topology optimization code written in matlab.
\newblock {\em Structural and multidisciplinary optimization}, 21:120--127, 2001.

\bibitem{xie1993simple}
Yi~Min Xie and Grant~P Steven.
\newblock A simple evolutionary procedure for structural optimization.
\newblock {\em Computers \& structures}, 49(5):885--896, 1993.

\bibitem{querin1998evolutionary}
Osvaldo~M Querin, Grant~P Steven, and Yi~Min Xie.
\newblock Evolutionary structural optimisation (eso) using a bidirectional algorithm.
\newblock {\em Engineering computations}, 15(8):1031--1048, 1998.

\bibitem{wang2003level}
Michael~Yu Wang, Xiaoming Wang, and Dongming Guo.
\newblock A level set method for structural topology optimization.
\newblock {\em Computer methods in applied mechanics and engineering}, 192(1-2):227--246, 2003.

\bibitem{allaire2004structural}
Gr{\'e}goire Allaire, Fran{\c{c}}ois Jouve, and Anca-Maria Toader.
\newblock Structural optimization using sensitivity analysis and a level-set method.
\newblock {\em Journal of computational physics}, 194(1):363--393, 2004.

\bibitem{mirzendehdel2015pareto}
Amir~M Mirzendehdel and Krishnan Suresh.
\newblock A pareto-optimal approach to multimaterial topology optimization.
\newblock {\em Journal of Mechanical Design}, 137(10):101701, 2015.

\bibitem{deng2015multi}
Shiguang Deng and Krishnan Suresh.
\newblock Multi-constrained topology optimization via the topological sensitivity.
\newblock {\em Structural and Multidisciplinary Optimization}, 51:987--1001, 2015.

\bibitem{du2019moving}
Bingxiao Du, Wen Yao, Yong Zhao, and Xiaoqian Chen.
\newblock A moving morphable voids approach for topology optimization with closed b-splines.
\newblock {\em Journal of Mechanical Design}, 141(8):081401, 2019.

\bibitem{guo2014topology}
Xu~Guo, Weisheng Zhang, and Wenliang Zhong.
\newblock Topology optimization based on moving deformable components: A new computational framework.
\newblock {\em arXiv preprint arXiv:1404.4820}, 2014.

\bibitem{zhang2017structural}
Weisheng Zhang, Wanying Yang, Jianhua Zhou, Dong Li, and Xu~Guo.
\newblock Structural topology optimization through explicit boundary evolution.
\newblock {\em Journal of Applied Mechanics}, 84(1):011011, 2017.

\bibitem{amir2012efficient}
Oded Amir, Ole Sigmund, Boyan~S Lazarov, and Mattias Schevenels.
\newblock Efficient reanalysis techniques for robust topology optimization.
\newblock {\em Computer Methods in Applied Mechanics and Engineering}, 245:217--231, 2012.

\bibitem{huang2017novel}
Guanxin Huang, Hu~Wang, and Guangyao Li.
\newblock A novel multi-grid assisted reanalysis for re-meshed finite element models.
\newblock {\em Computer Methods in Applied Mechanics and Engineering}, 313:817--833, 2017.

\bibitem{aage2015topology}
Niels Aage, Erik Andreassen, and Boyan~Stefanov Lazarov.
\newblock Topology optimization using petsc: An easy-to-use, fully parallel, open source topology optimization framework.
\newblock {\em Structural and Multidisciplinary Optimization}, 51:565--572, 2015.

\bibitem{shin2023topology}
Seungyeon Shin, Dongju Shin, and Namwoo Kang.
\newblock Topology optimization via machine learning and deep learning: A review.
\newblock {\em Journal of Computational Design and Engineering}, 10(4):1736--1766, 2023.

\bibitem{papadrakakis1998structural}
Manolis Papadrakakis, Nikos~D Lagaros, and Yiannis Tsompanakis.
\newblock Structural optimization using evolution strategies and neural networks.
\newblock {\em Computer methods in applied mechanics and engineering}, 156(1-4):309--333, 1998.

\bibitem{banga20183d}
Saurabh Banga, Harsh Gehani, Sanket Bhilare, Sagar Patel, and Levent Kara.
\newblock 3d topology optimization using convolutional neural networks.
\newblock {\em arXiv preprint arXiv:1808.07440}, 2018.

\bibitem{abueidda2020topology}
Diab~W Abueidda, Seid Koric, and Nahil~A Sobh.
\newblock Topology optimization of 2d structures with nonlinearities using deep learning.
\newblock {\em Computers \& Structures}, 237:106283, 2020.

\bibitem{kallioras2020accelerated}
Nikos~Ath Kallioras, Georgios Kazakis, and Nikos~D Lagaros.
\newblock Accelerated topology optimization by means of deep learning.
\newblock {\em Structural and Multidisciplinary Optimization}, 62(3):1185--1212, 2020.

\bibitem{li2019non}
Baotong Li, Congjia Huang, Xin Li, Shuai Zheng, and Jun Hong.
\newblock Non-iterative structural topology optimization using deep learning.
\newblock {\em Computer-Aided Design}, 115:172--180, 2019.

\bibitem{ulu2016data}
Erva Ulu, Rusheng Zhang, and Levent~Burak Kara.
\newblock A data-driven investigation and estimation of optimal topologies under variable loading configurations.
\newblock {\em Computer Methods in Biomechanics and Biomedical Engineering: Imaging \& Visualization}, 4(2):61--72, 2016.

\bibitem{nie2021topologygan}
Zhenguo Nie, Tong Lin, Haoliang Jiang, and Levent~Burak Kara.
\newblock Topologygan: Topology optimization using generative adversarial networks based on physical fields over the initial domain.
\newblock {\em Journal of Mechanical Design}, 143(3):031715, 2021.

\bibitem{sosnovik2019neural}
Ivan Sosnovik and Ivan Oseledets.
\newblock Neural networks for topology optimization.
\newblock {\em Russian Journal of Numerical Analysis and Mathematical Modelling}, 34(4):215--223, 2019.

\bibitem{yu2019deep}
Yonggyun Yu, Taeil Hur, Jaeho Jung, and In~Gwun Jang.
\newblock Deep learning for determining a near-optimal topological design without any iteration.
\newblock {\em Structural and Multidisciplinary Optimization}, 59(3):787--799, 2019.

\bibitem{zhou2020new}
Ying Zhou, Haifei Zhan, Weihong Zhang, Jihong Zhu, Jinshuai Bai, Qingxia Wang, and Yuantong Gu.
\newblock A new data-driven topology optimization framework for structural optimization.
\newblock {\em Computers \& Structures}, 239:106310, 2020.

\bibitem{zhang2021tonr}
Zeyu Zhang, Yu~Li, Weien Zhou, Xiaoqian Chen, Wen Yao, and Yong Zhao.
\newblock Tonr: An exploration for a novel way combining neural network with topology optimization.
\newblock {\em Computer Methods in Applied Mechanics and Engineering}, 386:114083, 2021.

\bibitem{deng2022self}
Changyu Deng, Yizhou Wang, Can Qin, Yun Fu, and Wei Lu.
\newblock Self-directed online machine learning for topology optimization.
\newblock {\em Nature communications}, 13(1):388, 2022.

\bibitem{chi2021universal}
Heng Chi, Yuyu Zhang, Tsz Ling~Elaine Tang, Lucia Mirabella, Livio Dalloro, Le~Song, and Glaucio~H Paulino.
\newblock Universal machine learning for topology optimization.
\newblock {\em Computer Methods in Applied Mechanics and Engineering}, 375:112739, 2021.

\bibitem{behzadi2021real}
Mohammad~Mahdi Behzadi and Horea~T Ilie{\c{s}}.
\newblock Real-time topology optimization in 3d via deep transfer learning.
\newblock {\em Computer-Aided Design}, 135:103014, 2021.

\bibitem{raissi2019physics}
Maziar Raissi, Paris Perdikaris, and George~E Karniadakis.
\newblock Physics-informed neural networks: A deep learning framework for solving forward and inverse problems involving nonlinear partial differential equations.
\newblock {\em Journal of Computational physics}, 378:686--707, 2019.

\bibitem{rao2021physics}
Chengping Rao, Hao Sun, and Yang Liu.
\newblock Physics-informed deep learning for computational elastodynamics without labeled data.
\newblock {\em Journal of Engineering Mechanics}, 147(8):04021043, 2021.

\bibitem{chen2021new}
Liang Chen and Mo-How~Herman Shen.
\newblock A new topology optimization approach by physics-informed deep learning process.
\newblock {\em Advances in Science, Technology and Engineering Systems Journal}, 6(4):233--240, 2021.

\bibitem{chandrasekhar2021tounn}
Aaditya Chandrasekhar and Krishnan Suresh.
\newblock Tounn: Topology optimization using neural networks.
\newblock {\em Structural and Multidisciplinary Optimization}, 63(3):1135--1149, 2021.

\bibitem{ji2023recent}
Wenye Ji, Jin Chang, He-Xiu Xu, Jian~Rong Gao, Simon Gr{\"o}blacher, H~Paul Urbach, and Aur{\`e}le~JL Adam.
\newblock Recent advances in metasurface design and quantum optics applications with machine learning, physics-informed neural networks, and topology optimization methods.
\newblock {\em Light: Science \& Applications}, 12(1):169, 2023.

\bibitem{yin2024dynamically}
Jichao Yin, Ziming Wen, Shuhao Li, Yaya Zhang, and Hu~Wang.
\newblock Dynamically configured physics-informed neural network in topology optimization applications.
\newblock {\em Computer Methods in Applied Mechanics and Engineering}, 426:117004, 2024.

\bibitem{li2021physics}
Wei Li, Martin~Z Bazant, and Juner Zhu.
\newblock A physics-guided neural network framework for elastic plates: Comparison of governing equations-based and energy-based approaches.
\newblock {\em Computer Methods in Applied Mechanics and Engineering}, 383:113933, 2021.

\bibitem{he2023deep}
Junyan He, Charul Chadha, Shashank Kushwaha, Seid Koric, Diab Abueidda, and Iwona Jasiuk.
\newblock Deep energy method in topology optimization applications.
\newblock {\em Acta Mechanica}, 234(4):1365--1379, 2023.

\bibitem{tijskens2002automatic}
Engelbert Tijskens, Dirk Roose, Herman Ramon, and Josse De~Baerdemaeker.
\newblock Automatic differentiation for solving nonlinear partial differential equations: an efficient operator overloading approach.
\newblock {\em Numerical Algorithms}, 30:259--301, 2002.

\bibitem{margossian2019review}
Charles~C Margossian.
\newblock A review of automatic differentiation and its efficient implementation.
\newblock {\em Wiley interdisciplinary reviews: data mining and knowledge discovery}, 9(4):e1305, 2019.

\bibitem{lu2021physics}
Lu~Lu, Raphael Pestourie, Wenjie Yao, Zhicheng Wang, Francesc Verdugo, and Steven~G Johnson.
\newblock Physics-informed neural networks with hard constraints for inverse design.
\newblock {\em SIAM Journal on Scientific Computing}, 43(6):B1105--B1132, 2021.

\bibitem{jeong2023physics}
Hyogu Jeong, Jinshuai Bai, Chanaka~Prabuddha Batuwatta-Gamage, Charith Rathnayaka, Ying Zhou, and YuanTong Gu.
\newblock A physics-informed neural network-based topology optimization (pinnto) framework for structural optimization.
\newblock {\em Engineering Structures}, 278:115484, 2023.

\bibitem{jeong2023complete}
Hyogu Jeong, Chanaka Batuwatta-Gamage, Jinshuai Bai, Yi~Min Xie, Charith Rathnayaka, Ying Zhou, and YuanTong Gu.
\newblock A complete physics-informed neural network-based framework for structural topology optimization.
\newblock {\em Computer Methods in Applied Mechanics and Engineering}, 417:116401, 2023.

\bibitem{joglekar2023dmf}
Aditya Joglekar, Hongrui Chen, and Levent~Burak Kara.
\newblock Dmf-tonn: direct mesh-free topology optimization using neural networks.
\newblock {\em Engineering with Computers}, pages 1--14, 2023.

\bibitem{zehnder2021ntopo}
Jonas Zehnder, Yue Li, Stelian Coros, and Bernhard Thomaszewski.
\newblock Ntopo: Mesh-free topology optimization using implicit neural representations.
\newblock {\em Advances in Neural Information Processing Systems}, 34:10368--10381, 2021.

\bibitem{kumar2021topology}
P~Kumar, C~Schmidleithner, NB~Larsen, and O~Sigmund.
\newblock Topology optimization and 3d printing of large deformation compliant mechanisms for straining biological tissues.
\newblock {\em Structural and Multidisciplinary Optimization}, 63:1351--1366, 2021.

\bibitem{sitzmann2020implicit}
Vincent Sitzmann, Julien Martel, Alexander Bergman, David Lindell, and Gordon Wetzstein.
\newblock Implicit neural representations with periodic activation functions.
\newblock {\em Advances in neural information processing systems}, 33:7462--7473, 2020.

\bibitem{samaniego2020energy}
Esteban Samaniego, Cosmin Anitescu, Somdatta Goswami, Vien~Minh Nguyen-Thanh, Hongwei Guo, Khader Hamdia, Xiaoying Zhuang, and Timon Rabczuk.
\newblock An energy approach to the solution of partial differential equations in computational mechanics via machine learning: Concepts, implementation and applications.
\newblock {\em Computer Methods in Applied Mechanics and Engineering}, 362:112790, 2020.

\bibitem{goswami2020adaptive}
Somdatta Goswami, Cosmin Anitescu, and Timon Rabczuk.
\newblock Adaptive fourth-order phase field analysis using deep energy minimization.
\newblock {\em Theoretical and Applied Fracture Mechanics}, 107:102527, 2020.

\bibitem{wang2023dcem}
Yizheng Wang, Jia Sun, Timon Rabczuk, and Yinghua Liu.
\newblock Dcem: A deep complementary energy method for linear elasticity.
\newblock {\em International Journal for Numerical Methods in Engineering}, page e7585, 2023.

\bibitem{goswami2020transfer}
Somdatta Goswami, Cosmin Anitescu, Souvik Chakraborty, and Timon Rabczuk.
\newblock Transfer learning enhanced physics informed neural network for phase-field modeling of fracture.
\newblock {\em Theoretical and Applied Fracture Mechanics}, 106:102447, 2020.

\bibitem{white2018toplogical}
Daniel~A White, Mark~L Stowell, and Daniel~A Tortorelli.
\newblock Toplogical optimization of structures using fourier representations.
\newblock {\em Structural and Multidisciplinary Optimization}, 58:1205--1220, 2018.

\bibitem{tancik2020fourier}
Matthew Tancik, Pratul Srinivasan, Ben Mildenhall, Sara Fridovich-Keil, Nithin Raghavan, Utkarsh Singhal, Ravi Ramamoorthi, Jonathan Barron, and Ren Ng.
\newblock Fourier features let networks learn high frequency functions in low dimensional domains.
\newblock {\em Advances in neural information processing systems}, 33:7537--7547, 2020.

\bibitem{doosti2021topology}
Nikan Doosti, Julian Panetta, and Vahid Babaei.
\newblock Topology optimization via frequency tuning of neural design representations.
\newblock In {\em Proceedings of the 6th Annual ACM Symposium on Computational Fabrication}, pages 1--9, 2021.

\bibitem{nguyen2020deep}
Vien~Minh Nguyen-Thanh, Xiaoying Zhuang, and Timon Rabczuk.
\newblock A deep energy method for finite deformation hyperelasticity.
\newblock {\em European Journal of Mechanics-A/Solids}, 80:103874, 2020.

\bibitem{andreassen2011efficient}
Erik Andreassen, Anders Clausen, Mattias Schevenels, Boyan~S Lazarov, and Ole Sigmund.
\newblock Efficient topology optimization in matlab using 88 lines of code.
\newblock {\em Structural and Multidisciplinary Optimization}, 43:1--16, 2011.

\end{thebibliography}
\end{document}